\documentclass[aps,pra,letterpaper,10pt,twocolumn,showpacs,superscriptaddress,notitlepage]{revtex4-1}

\usepackage[utf8]{inputenc}
\usepackage{amsfonts,amssymb,amsmath,amsthm}
\usepackage{url}
\usepackage{dsfont}
\usepackage{bbm}
\usepackage[usenames,dvipsnames]{xcolor}
\usepackage{xcolor}
\usepackage{nicefrac}
\usepackage{natbib}
\usepackage{setspace}

\usepackage{tabularx,booktabs}
\usepackage{comment}
\usepackage[normalem]{ulem}
\usepackage{todonotes}
\usepackage{siunitx}
\usepackage{comment}
\usepackage{optidef}
\usepackage{graphicx}

\usepackage[colorlinks=true, citecolor=blue,urlcolor=blue,linkcolor=blue,filecolor=black]{hyperref}

\def\bra #1{\langle #1\vert}
\def\ket #1{\vert #1\rangle}

\newcommand{\beq}{\begin{equation}}
\newcommand{\eeq}{\end{equation}}

\newcommand{\Hmin}{H_{\rm min}(X|E)}
\newcommand{\pguess}{p_{\rm g}}
\newcommand{\Id}{\hat{\mathds{1}}}

\newtheorem{theorem}{Theorem}

\newtheorem{prop}[theorem]{Proposition}

\DeclareMathOperator{\Tr}{Tr}

\definecolor{marcocol}{rgb}{.1,0.6,0.9}

\definecolor{pinocol}{rgb}{.1,0.7,0.1}

\newcommand{\titlename}{Unbounded randomness from uncharacterized sources}
\begin{document}

\title{\titlename}

 \author{Marco Avesani}
 \affiliation{Dipartimento di Ingegneria dell'Informazione, Universit\`a degli Studi di Padova, via Gradenigo 6B, 35131 Padova, Italia}

 \author{Hamid Tebyanian}
 \affiliation{Dipartimento di Ingegneria dell'Informazione, Universit\`a degli Studi di Padova, via Gradenigo 6B, 35131 Padova, Italia}

 \author{Paolo Villoresi}
 \affiliation{Dipartimento di Ingegneria dell'Informazione, Universit\`a degli Studi di Padova, via Gradenigo 6B, 35131 Padova, Italia}
\affiliation{Istituto di Fotonica e Nanotecnologie - CNR, Via Trasea 7, 35131 Padova, Italia}
 
 \author{Giuseppe Vallone}
\affiliation{Dipartimento di Ingegneria dell'Informazione, Universit\`a degli Studi di Padova, via Gradenigo 6B, 35131 Padova, Italia}
\affiliation{Dipartimento di Fisica e Astronomia, Universit\`a di Padova, via Marzolo 8, 35131 Padova, Italia}
 \affiliation{Istituto di Fotonica e Nanotecnologie - CNR, Via Trasea 7, 35131 Padova, Italia}

%%%%%%%%%%%%%%%%%%%%%%%%%%%%%%%%%%%%%%%%%%%%%%%%%%%%%%%%%%%%%%%%%%%%%%%%%%%

%%%%%%%%%%%%%%%%%%%%%%%%%%%%%%%%%%%%%%%%%%%%%%%%%%%%%%%%%%%%%%%%%%%%%%%%%%%

%%%%%%%%%%%%%%%%%%%%%%%%%%%%%%%%%%%%%%%%%%%%%%%%%%%%%%%%%%%%%%%%%%%%%%%%%%%

\begin{abstract}
Randomness is a central feature of quantum mechanics and an invaluable resource for both classical and quantum technologies.
Commonly, in Device-Independent and Semi-Device-Independent scenarios, randomness is certified using projective measurements and the amount of certified randomness is bounded by the dimension of the measured quantum system.
In this work, we propose a new Source-Device-Independent protocol, based on Positive Operator Valued Measurement (POVM), which can arbitrarily increase the number of certified bits for any fixed dimension.
A tight lower-bound on the quantum conditional min-entropy is derived using only the POVM structure and the experimental expectation values, taking into account the quantum side-information.  For symmetrical POVM measurements on the Bloch sphere we have derived closed-form analytical bounds. 
Finally, we experimentally demonstrate our method with a compact and simple  photonic setup that employs polarization-encoded qubits and POVM up to 6 outcomes.

\end{abstract}

\maketitle

\section{Introduction}

Random numbers are necessary for many different applications, ranging from simulations to cryptography and tests of fundamental physics, such as Bell tests~\cite{Hensen2015,Giustina2015,Shalm2015}.
Despite their common use, the  certification of randomness is a complex task. 
Classical processes cannot generate genuine randomness due to the determinism of classical mechanics.
On the other hand, randomness is an intrinsic feature of quantum mechanics due to the probabilistic nature of its laws.
However, the generation and certification of randomness, even from quantum processes, always requires some assumptions \cite{Acin2016}.

The most reliable type of certification is given by \textit{Device-Independent} (DI) protocols \cite{Acin2016} where the violation of a Bell inequality can certify the randomness and the privacy of the numbers without any assumption on the devices used. Despite recent demonstrations \cite{Bierhorst2017,Liu2018,Liu2019b,Zhang2020,Shalm2019}, DI-QRNGs are extremely demanding from the experimental point of view and also their performances cannot satisfy the needs of practical implementations. For this reason, all current commercial QRNGs use \textit{trusted protocols}, where both the source and the measurements are trusted. 

Even though trusted QRNGs are high-rate, easy-to-implement, and cheap,  the security and the privacy of the generated random numbers could be compromised.
Recently a new class of protocols, called \textit{Semi-Device-Independent} (Semi-DI) \cite{Herrero-Collantes2017,Ma2016} have been proposed as a compromise between the DI and the trusted ones. The Semi-DI protocols work in the similarly ``paranoid scenario'' of DI, although with few assumptions on the devices' inner working. The assumptions can be related to the dimension of the exchanged system\cite{Lunghi2014}, the source \cite{Vallone2014a,Marangon2017,Avesani2018,Drahi2019,Smith2019}, the measurement \cite{Nie2016,Bischof2017}, the overlap between the states \cite{Brask2017} or the energy \cite{Rusca2019,Himbeeck2017semidevice,VanHimbeeck2019,Avesani2020,rusca2020,tebyanian2020semidevice}. These protocols are promising, since they can provide a higher level of security with a generation rate compatible with practical needs.

Most of the DI and Semi-DI protocols employ projective measurement, limiting the maximal certification to the underlying Hilbert space's dimension.  
The possibility to increase the generation rate using general measurement has been recently discussed for entangled systems in the DI scenario \cite{Acin2016a,Andersson2018,Gomez2017}.
While projective measurements can only certify up to one bit of randomness for every pair of entangled qubits, POVM can saturate the optimal bound of 2 bits \cite{Andersson2018}. Additionally, unbounded generation is possible if repeated non-demolition measurements are performed on one of the qubits, but the protocol is not robust to noise~\cite{Curchod2017}.
Yet, all these scenarios need entanglement, which is a strong requirement and involves an increased experimental complexity. 

In this work, we will consider a prepare-and-measure scenario where the coherence (or purity) of the source is the resource for the protocol. 
We will show that a robust unbounded randomness certification can be obtained in the Source-DI scenario when non-orthogonal POVMs are used. 
For a fixed dimension of Hilbert space, we demonstrate that the amount of extractable random bits scales up as $\propto \log_2(N)$ with $N$ the number of POVM outcomes.
In such a way, an infinite number of random bits can be certified for any dimension of the quantum system to be measured.
We specialize our analysis for polarization qubits, considering symmetric POVM measurements. In particular, we derive tight analytical bounds for equiangular POVMs restricted on the plane of the Bloch sphere and for POVMs that correspond to Platonic solids inscribed in the Bloch sphere. 
Finally, to validate our findings, we experimentally implement  three equiangular measurements on a plane with $3$, $4$ and $6$ outcomes, and the octahedron measurement with $6$ outcomes, using a simple optical setup. 
\section{Randomness certification with POVM}
In the prepare and measure scenario a QRNG is composed of two systems: a source, that emits a quantum state $\hat \rho_A$ and a measurement station. At each round, the measurement produces an outcome $K=k$ with some probability $P_k$.

While in the trusted scenario, both the measurement and preparation stages are trusted and characterized, in the Source-DI scenario only the measurement is characterized, while the source is considered untrusted and under the control of the eavesdropper(Eve). 
In this case the amount of \textbf{private} randomness that can be extracted by the QRNG can be quantified by quantum conditional min-entropy~\cite{Tomamichel2011},
related to the guessing probability as
\beq
\label{Hmin_pg}
\Hmin=-\log_2\left(\pguess(X|E)\right)
\eeq
Here, the probability of correctly
guessing the measurement outcome $p_g$ is conditioned on Eve's (quantum) side information $E$ on the system. 

As discussed in \cite{Vallone2014a}, if the prepared state $\hat \rho_A$ is pure, Eve does not have access to any quantum side information.
On the contrary, if $\hat \rho_A$ is mixed, there always exists a purification $\hat \rho_{AE}$ of $\hat \rho_{A}$, such that the systems $A$ and $E$ are correlated.
Bounding the $H_{min}(X|E)$ is then directly linked with the problem of bounding the purity of the unknown state $\hat \rho_A$.

In this scenario, a single projective measurement $\{\hat P_i\}$ cannot certify any amount of randomness~\cite{Fiorentino2007a,Vallone2014a}.

A solution to this problem, proposed in  \cite{Vallone2014a}, uses two conjugate projective measurements $\mathds{Z}$ and $\mathds{X}$
and the Entropic Uncertainty Principle to bound the value of $\Hmin$.
However, this approach requires the active switching of the two conjugate measurements that comes with two major drawbacks: first, the switching requires an initial source of private randomness and then requires active elements in the experimental implementation, increasing the complexity of the setup. 
For this protocol, the maximal value of min-entropy is upper bounded by the dimension $d$ of the measurement $H_{min}(X|E)\leq \log_2(d)$.

In the following, we will show that the use of a single POVM $\{ \hat F_k \}$ with $k=1,\cdots,N$ at the measurement station will solve the above issues. No initial randomness and no active devices are required; the maximal value of the min-entropy is bounded by the number of POVM elements $N$, but is not limited by the dimension $d$ of the underlying Hilbert space.

As shown in~\cite{InPreparation,Avesani2018}, in this scenario the guessing probability can be written as
\beq
\pguess(X|E)=  \max_{\{\hat\tau_k \}} \sum_{k=1}^N\Tr_{A}\left[ \hat F_k  \hat \tau_k\right]\,
\label{eq:pguess_SDI}
\eeq
with the following constraint on the 
sub-normalized states $\hat \tau_x$
\beq
\Tr[\hat F_j(\sum_k\hat\tau_k)]=P_j \,, \quad 
j=1,\cdots, N\,.
\eeq
The above constraint ensures that the states $\hat \tau_k$ form a decomposition of the state $\hat \rho'_A\equiv \sum_k\hat\tau_k$ that has the same outcome probabilities $P_j$ of the unknown state $\hat\rho_A$ when measured with the POVM $\{\hat F_k\}$. Differently from the the common definition of $\pguess$ for classical-quantum states presented in \cite{Konig2009}, the above formulation is easier to calculate in scenarios where the source is untrusted (see~\cite{InPreparation} for more details).

From this definition of the $\pguess(X|E)$, we can see why a single projective measurement $\{\hat \Pi_k\}$ (that satisfies $\hat \Pi_j\hat \Pi_k=\delta_{j,k}\hat \Pi_k$) cannot be used to extract randomness: for every set of $P_x$ we can choose in Eq. \ref{eq:pguess_SDI} $\hat \tau_k = P_k \hat \Pi_k$, such that $\sum_j\Tr[\hat \Pi_k \tau_j]=P_k$ and $\Tr[\hat \Pi_k \tau_k]=P_k$.
Thus, $\pguess(X|E)$ reaches unity, meaning that Eve is able to guess Alice's result deterministically.
On the other hand, if the POVM used by Alice have non-orthogonal elements $\hat F^N_j \hat F^N_k \neq \delta_{j,k}$ the attacker can never guess with certainty the outcome of the measurement.

Let's first consider the simple case where the equiangular three-state POVM for a qubit is used, namely:
\begin{align}
\label{3Pi}
\hat F_k^3&=\frac{2}{3}\ket{\psi_k}\bra{\psi_k}
\end{align}
where
\begin{equation}
\label{eq:psi_k}
\begin{aligned}
\ket{\psi_1}&=\ket{0}\\
\ket{\psi_2}&=\frac{1}{2}\ket{0}+\frac{\sqrt{3}}{2}\ket{1} \\
\ket{\psi_3}&=\frac{1}{2}\ket{0}-\frac{\sqrt{3}}{2}\ket{1}
\end{aligned}
\end{equation}
By solving the optimization problem in Eq. \ref{eq:pguess_SDI}, we can calculate the $\Hmin$ for every possible set of states $\hat \rho_A$ sent by the attacker. 

\begin{figure}[t]
{\centering
\includegraphics[width=\linewidth]{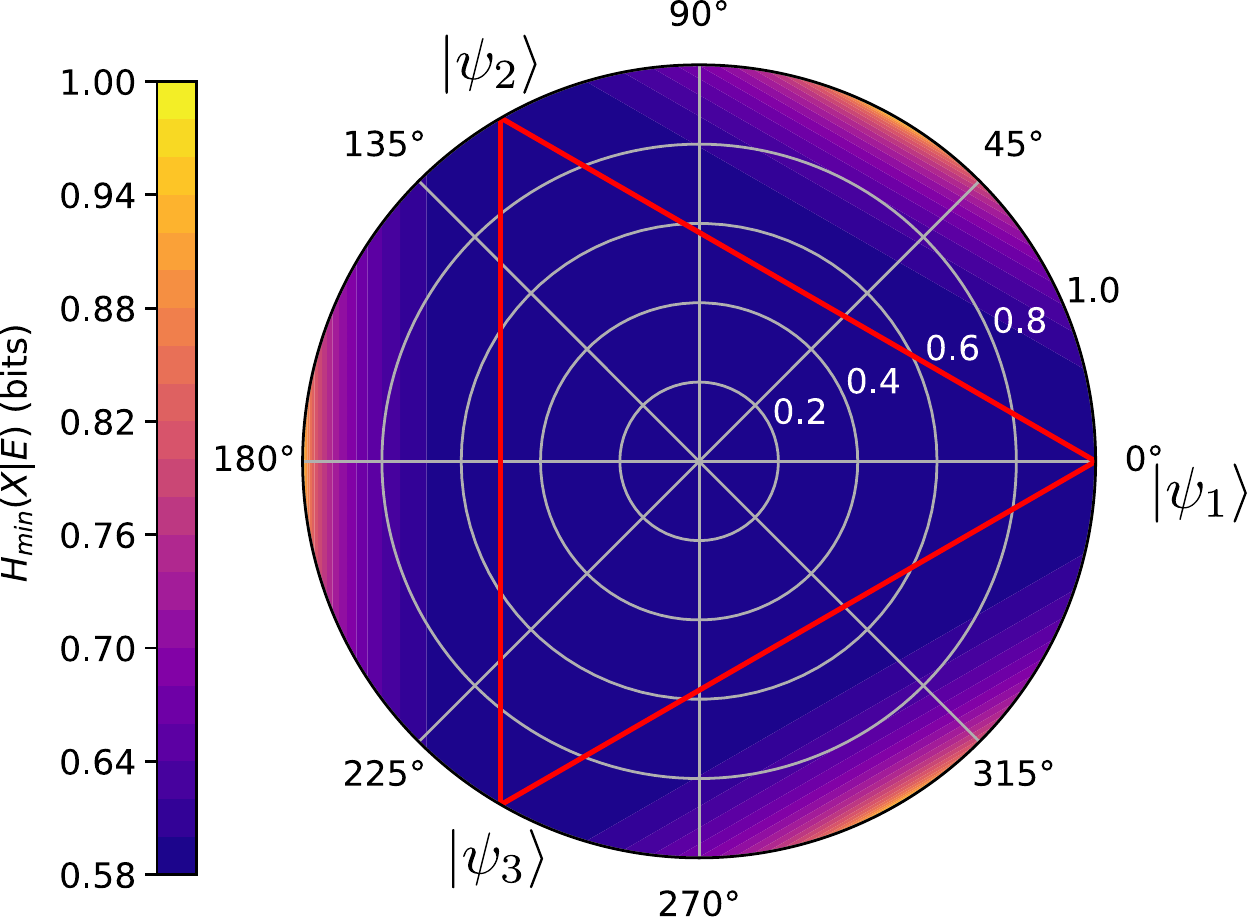}}
\caption{Contour plot of $H_{min}(X|E)$ for the three-outcome POVM $\hat F_k^3$ in function of the projection of $\hat \rho_A$ into the $ZX$ plane of the Bloch sphere. By this measurement it is always possible to certify more than 0.58 bits of randomness per measurement.}
\label{fig:3d_hmin}
\end{figure}

Since the POVM elements $\hat F_k$ belong to the $ZX$ plane of the Bloch sphere, all the $\hat \rho_A$ that have the same projection in the $ZX$ plane will lead to the same result.
The min-entropy in function of the projection of the state $\hat\rho_A$ in the  $ZX$ plane are shown in Fig. \ref{fig:3d_hmin}.

It is possible to distinguish two different areas: the region inside the triangle (formed by the lines that connect the three $\ket{\psi_k}$), and the one outside it. Inside this region, the min-entropy is constant and it reaches the minimal value of  $H_{min}(X|E)=-\log_2\left(2/3\right)\approx0.58$. This result is in contrast with projective measurements, where a single projective measurement can never achieve $H_{min}(X|E) >0$.
Outside this region, the min-entropy monotonically increases and reaches its maximum $H_{min}(X|E) =1$ for three pure states, each orthogonal to one of the states $\ket{\psi_k}$

The reason can be intuitively understood. Consider that the state orthogonal to $\ket{\psi_1}$ is sent: the output corresponding to $\hat F^{3}_1$ never appears and this result alone certifies the purity of $\rho_A$. On the other hand, the other two outcomes relative to $\hat F^3_2$ and $\hat F^3_3$ happen with equal probability of $0.5$. Then, in this case, it behaves like an unbiased coin, and the maximum achievable randomness is $1$ bit per measurement.

Exploiting the geometrical properties of the POVM we derived an analytical relation on $\pguess(X|E)$ as a function of the measured outcomes for general regular POVMs with $N$ outcomes, as stated below.

{\prop Consider the $N$-outcome qubit POVM $\{ \hat F^N_k \}$ defined by
\beq
\hat F^N_k =\frac13(\openone+\vec a_k\cdot \vec \sigma)\,,
\quad k=1,\cdots,N
\eeq
with $\vec a_k$ representing the vertices of a regular polygon in the $ZX$ plane, namely $\vec a_k=1$ and $\vec a_k\cdot\vec a_{k+1}=\cos\frac{2\pi}{N}$.
The measured output probabilities $P_k$ uniquely identify a point $R$ in the $ZX$ plane with coordinates $(r_z,r_x)$. 
The guessing probability $\pguess(X|E)$ is given by 

\beq
\begin{aligned}
\label{eq:Pg_teo}
p_g&=
\frac1N+\frac1N\sum_k
f_N(\vec r\cdot \vec u_k,\alpha)
\,
\theta(\vec r\cdot \vec u_k-\cos\alpha)
\end{aligned}
\eeq
where
\beq
\begin{aligned}
f_N(x,\alpha)&=
x\cos\alpha+\sqrt{1-x^2}\sin\alpha\\
\alpha &= \cos{\frac{\pi}{N}}
\end{aligned}
\eeq
and $\vec u_k$ the unit vectors orthogonal to the polygon edges.
}

We note that if the point $R$ is inside the polygon then $\pguess(X|E)=2/N$. Otherwise, if the point $R$ is outside the polygon, only one term in the sum \eqref{eq:Pg_teo} is nonvanishing.

\begin{figure}[t]
    \centering
    \includegraphics[width=0.49\linewidth]{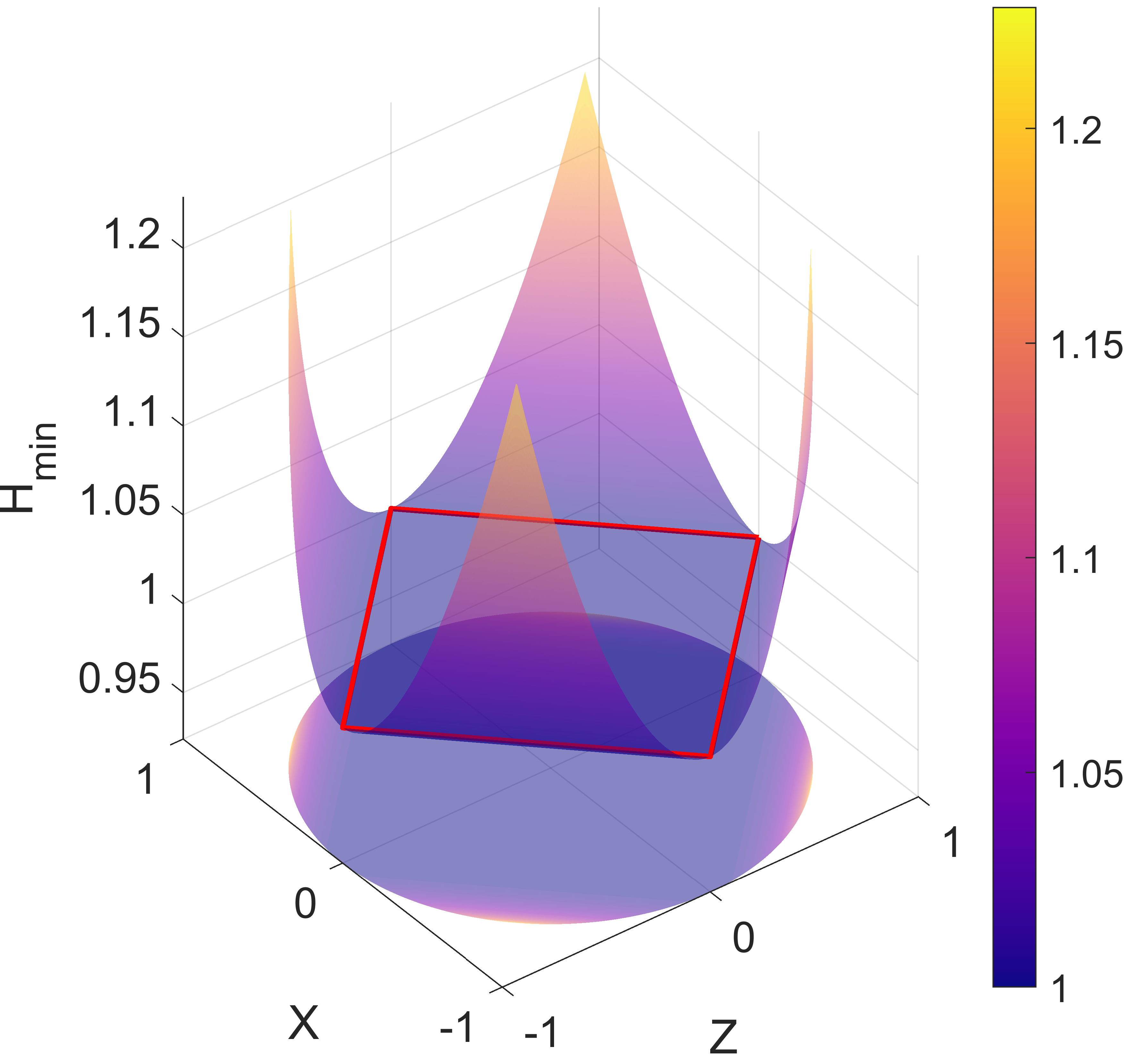}
    %\quad
    \includegraphics[width=0.49\linewidth]{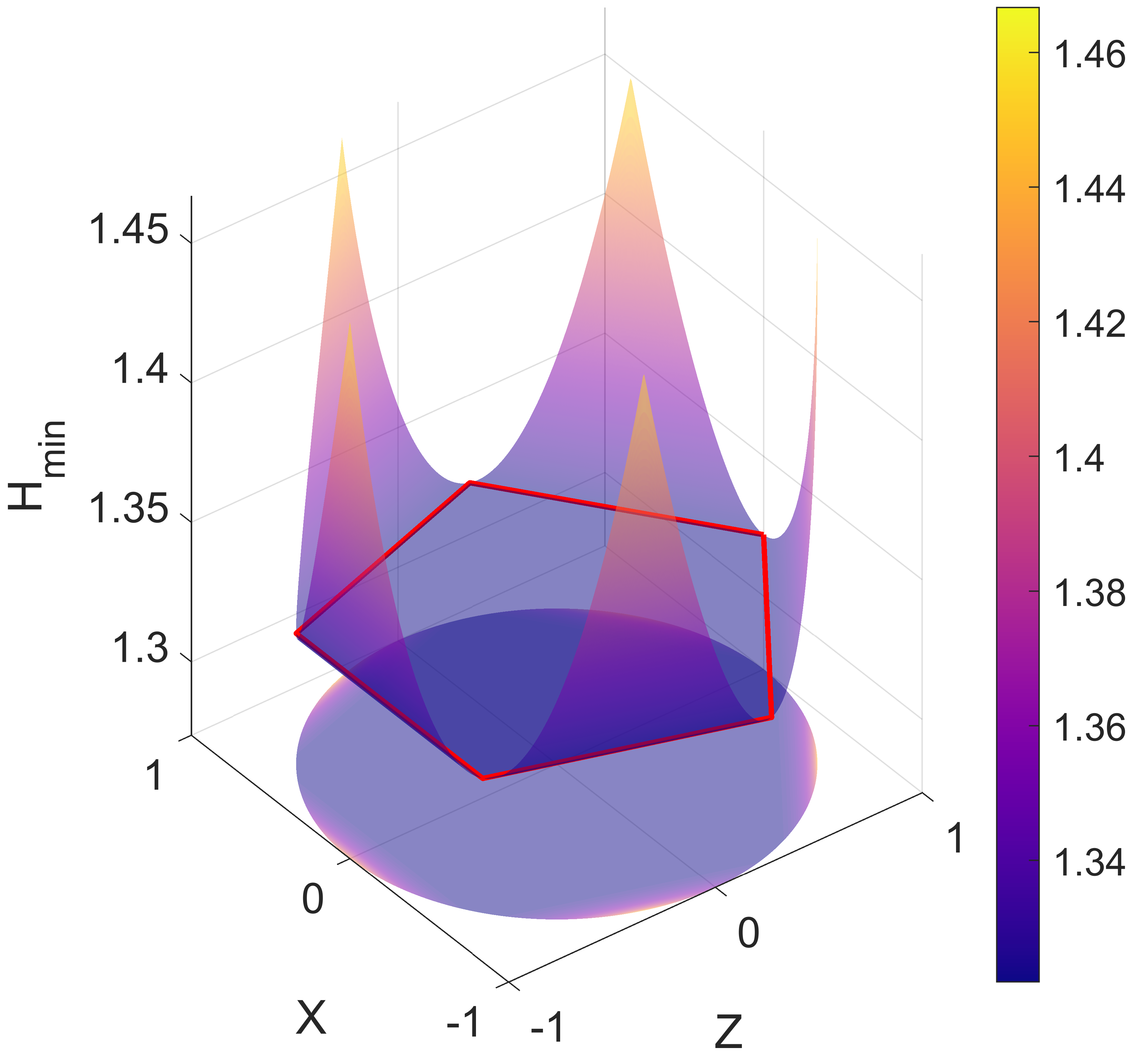}
    \\
    \includegraphics[width=0.49\linewidth]{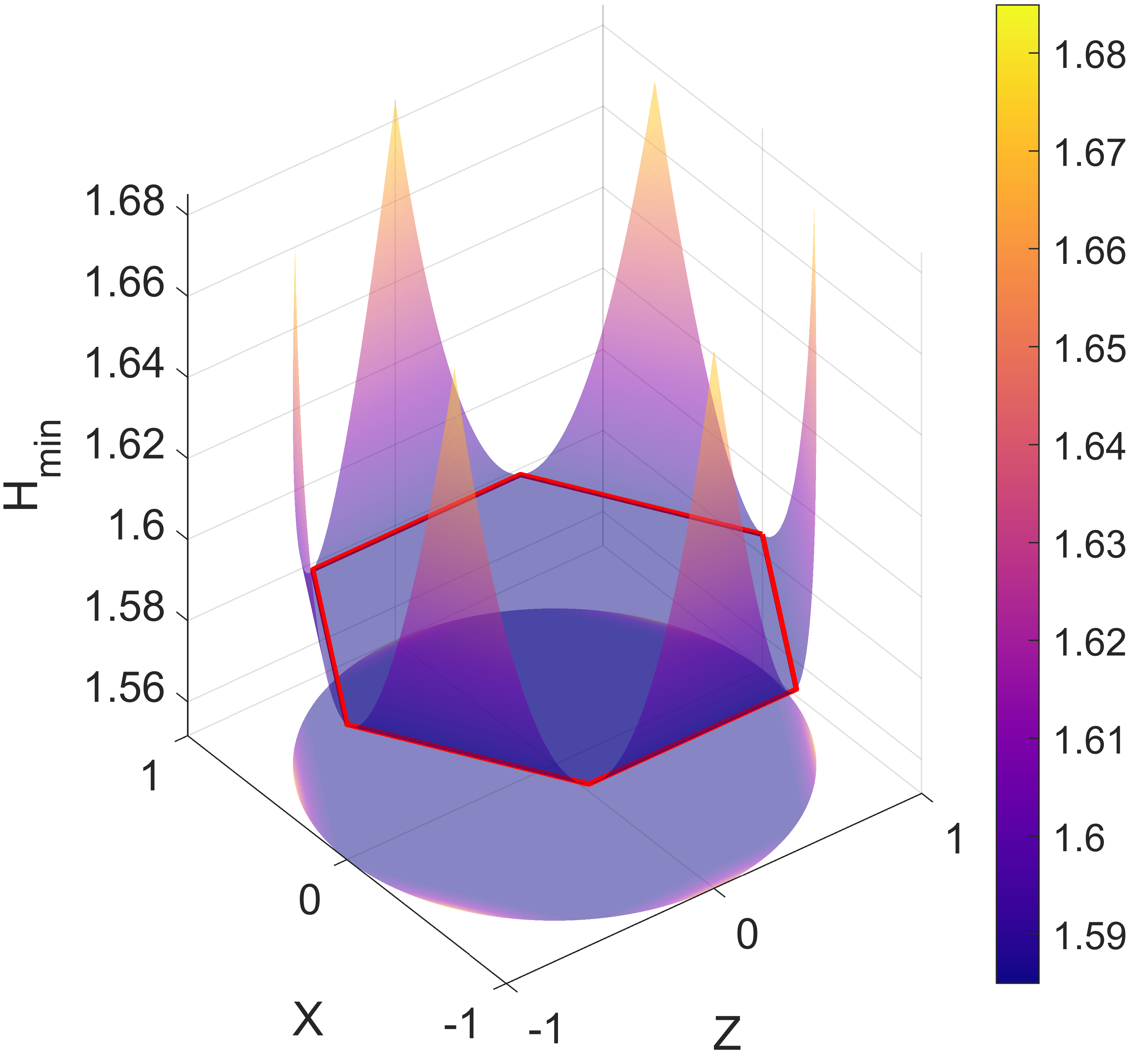}
    %\quad
    \includegraphics[width=0.49\linewidth]{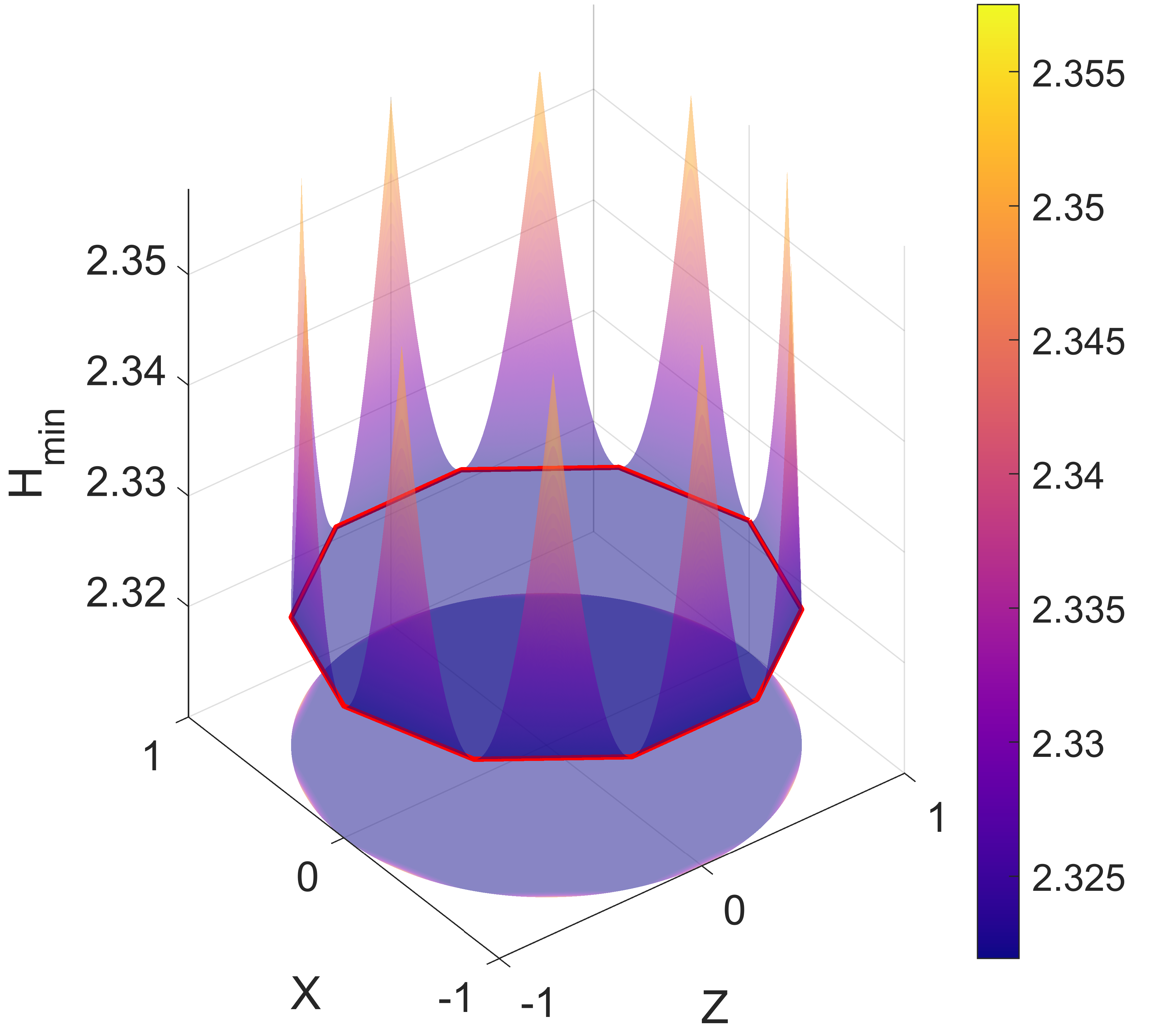}
    \caption{3D  plot of $H_{min}(X|E)$ in function of the projection of $\hat \rho_A$ in the $ZX$ plane. The POVM considered have $4,5,6,10$ equispaced elements in the $ZX$ plane.}
    \label{fig:TS_Hmin}
\end{figure}

The analytical results has been compared with the numerical solutions of Eq. \ref{eq:Pg_teo} for $N$ up to 100. 

The results were calculated respect the statistics reproduced by $\hat\rho_A$ sampled from the entire Bloch sphere. The numerical  and analytical method always agreed, up to a factor smaller than the numerical tolerance.

\begin{figure}[t]
    \includegraphics[width=\linewidth]{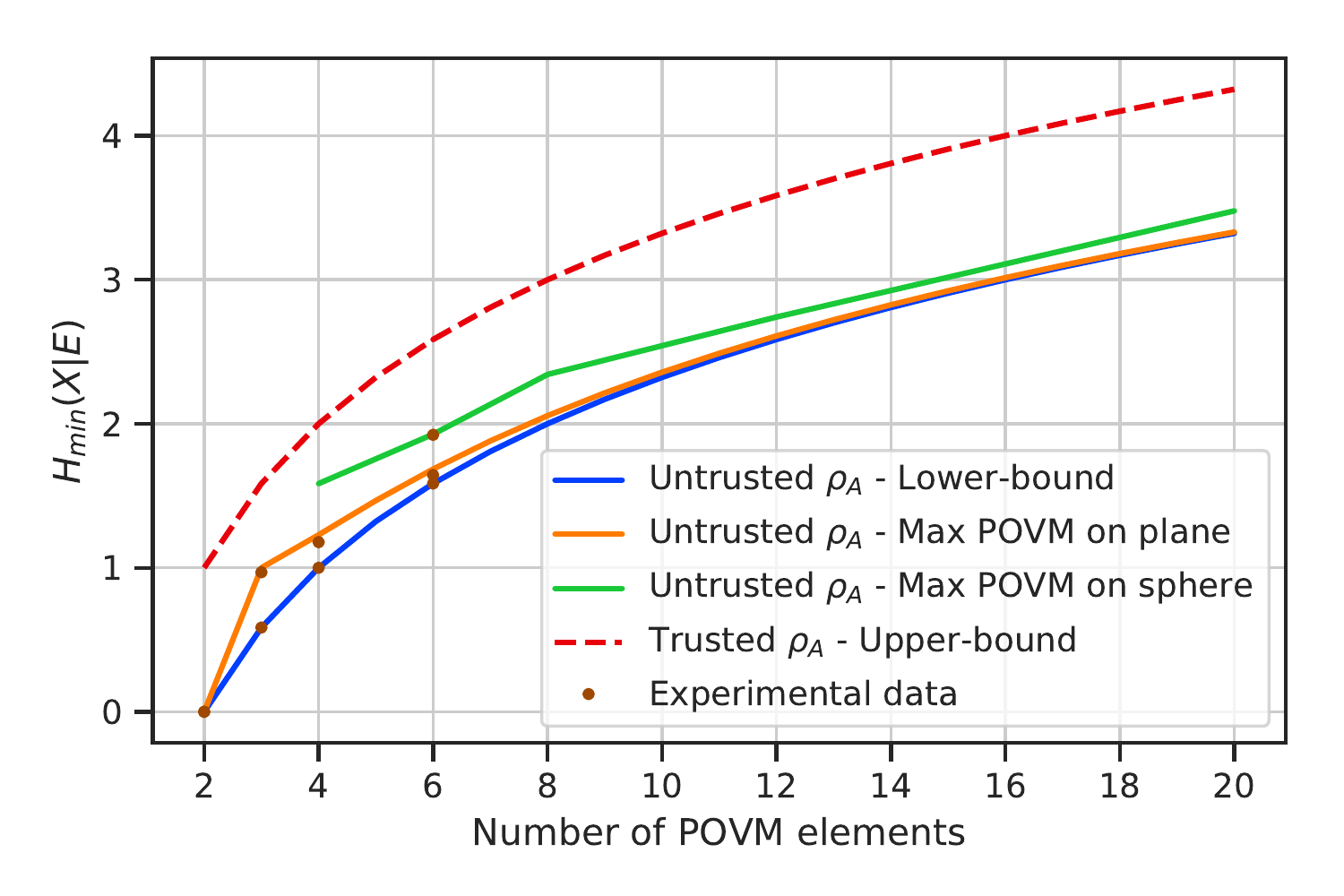}
    \caption{Scaling of $H_{min}(X|E)$ as a function of $N$. In the untrusted scenario, we report the max and the min $ H_{min}(X|E)$ for $N$ equispaced POVM on a plane of the Bloch sphere. We also report the max $H_{min}(X|E)$ for POVM representing platonic solids inscribed in the Bloch sphere. The dashed line represents the upper-bound for the trusted model. Finally, the colored dot represent estimated min-entropy from the experimental data.}
    \label{fig:POVM_scaling}
\end{figure}

In Fig. \ref{fig:TS_Hmin}, for $N=4,\,5,\,6$ and $10$, we show the contour plots of the min-entropy as a function of the projection 
of the unknown state $\hat \rho_A$ in the $ZX$ plane of the Bloch sphere.
By increasing the number of outcomes both the lowest and the highest $H_{min}(X|E)$ increase. From Eq.~\ref{eq:Pg_teo} we obtain:
\begin{align}
\label{eq:hmin_max}
    M_N\equiv\max_{\vec r} \left( H_{min}(X|E) \right) &= \log_2(\frac{N}{1+\cos{\alpha}} ) \\
    m_N\equiv\min_{\vec r} \left( H_{min}(X|E) \right)&= \log_2\left( N \right) - 1 
\end{align}
with $\alpha=\frac{\pi}{N}$.
This scaling as a function of $N$ for a qubit system and equiangular POVM on a plane is reported in Fig. \ref{fig:POVM_scaling}.

The difference between $M_N$ and $m_N$ is given by:
\beq
M_N-m_N= 1-\log_2\left( 1+\cos{\frac{\pi}{N}} \right) \approx \frac{\pi^2}{2N^2\ln{2}}
\eeq
which becomes negligible for large $N$, since the distance between the POVM's elements also gets smaller.

The analytical bounds of Eq. \ref{eq:Pg_teo} can be extended to general POVM, not restricted to a plane of the Bloch sphere. We also considered symmetric POVMs, representing platonic solids inscribed in the Bloch sphere \cite{Somczynski2016}.
We show in Appendix \ref{app:nequi} that for these measurements,  Eq. \ref{eq:hmin_max}, with different values of $\alpha$, correctly bounds the maximum amount of min-entropy that can be certified. In Fig. \ref{fig:POVM_scaling} we compare the scaling of such measurements with the POVM restricted to the plane.

Additionally, in Fig.~\ref{fig:POVM_scaling} we show a comparison between the extractable randomness in the trusted and in the Source-DI scenarios. In the trusted scenario (i.e. both source and measurement trusted, without quantum correlation between the devices and the attacker), up to $\log_2N$ bits can be certified per measurement, sending for example the completely mixed state $\Id_2$.

The gap between the trusted and the unstrusted bounds are never larger than 1 bit, for any $\rho_A$, meaning that the price to pay for the increased security of the Source-DI certification is at most $1$ bit per measurement.
Finally, the results indicate that in the asymptotic limit $N\rightarrow\infty$ the min-entropy tends to $H_{min}(X|E) \rightarrow\infty$, showing that unbounded randomness can be certified even from quantum systems with finite dimension $d$, including qubits.

\section{Experimental implementation}

\begin{figure*}[!htbp]
\centering
\includegraphics[width=0.75\linewidth]{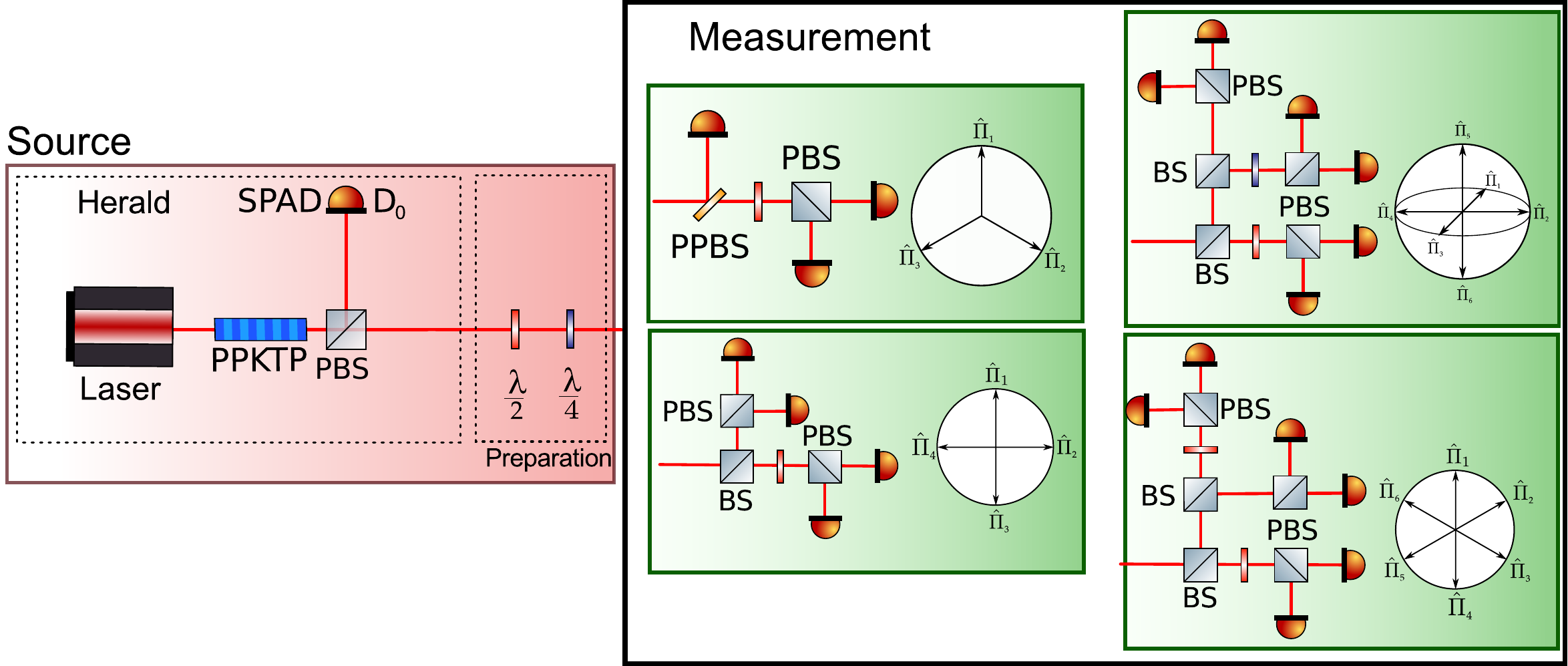}
\caption{An heralded single photon source generates single photons at 808 nm. After the polarizing beam splitter (PBS), the heralded photon is prepared 
in any desired polarization by using an Half-Wave Plate (HWP) and a Quarter-Wave Plate (QWP). The photon is then measured with four different POVM configurations with $3,4$ and $6$ outcomes. The coincidences between the heralding photon and the detectors in the measurement station are recorded by a timetagger on a PC.
} 
\label{fig:setup}
\end{figure*}

In order to experimentally test the certification protocol, we developed a simple optical setup that employs a heralded single-photon source and four different POVM configurations. The preparation and measurement exploit the polarization degree of freedom of single photons. A schematic representation of the setup is shown in Fig. \ref{fig:setup}.
The heralded source is composed of a continuous-wave (CW) laser at 404nm, which optically pumps a 30mm long Periodically Poled Potassium Titanyl Phosphate (PPKTP) crystal.
This configuration produces photon pairs at 808 nm through type-II collinear-phase-matching spontaneous parametric down-conversion (SPDC). 

The photons are deterministically separated by a polarizing beam splitter (PBS), and the detection of a photon at $D_0$ (see Fig. \ref{fig:setup}) heralds the presence of the single photon $\ket{H}_s$, which is sent to the preparation stage.

Here a Half Wave Plate (HWP) and a Quarter Wave Plate (QWP) are used to prepare the photon in any required polarization. The photon is then sent to Alice's measurement. Taking into account filtering and finite SPAD efficiency, we obtain a heralded photon generation rate of $\approx 10 \si{\kilo\hertz}$. 

We decided to implement the protocol using a heralded single-photon source in order to reduce the contribution of dark-counts and background noise. However, since we work in the Source-DI scenario, no assumptions are made on the source and any implementation can be used.

The POVM $\{\hat F^N_k\}$ used by Alice are $N$-output measurement in the two dimensional Hilbert space of photon polarization. 
The optical implementation of such POVM can be realized by using interferometric setups (as in \cite{Clarke2001}): however, this technique requires high precision in the alignment and offers low temporal stability. For this reason, we decided to follow the approach presented in~\cite{Schiavon2016}, which does not require any interferometric scheme.

In the three outcomes equiangular POVM $\hat F^3_k$, shown in Fig. \ref{fig:setup}, the photon passes through a Partially Polarizing Beam Splitter (pPBS), that reflects with probability $2/3$ the state $\ket{V}$, while fully transmits $\ket{H}$.

Thus, detecting the reflected photons implements the first POVM element $\hat F_1^3=\frac{2}{3}\ket{V}\bra{V}$. 
The transmitted part is instead measured in the diagonal basis, implementing the remaining operators $\hat F_2^3$ and $\hat F_3^3$ (see~\cite{Schiavon2016} for more detail).

The POVM with four and six outcomes can be implemented in a similar way and they only require standard BS, PBS and waveplates.
The four-outcome POVM $\hat F^4_k$ is realized in the following way: a $50:50$ BS reflects and transmits the photons with equal probability, then in the reflected path a PBS measures in the $\mathds{Z}$ basis, while in the transmitted path the HWP at $\frac{\pi}{8}$ followed by the PBS, performs a measurement in the $\mathds{X}$ basis.
Accordingly, the four POVM elements $\{\hat F^4_k\}=\{ \frac{1}{4}\ket{H}\bra{H},\frac{1}{4}\ket{+}\bra{+},\frac{1}{4}\ket{V}\bra{V},\frac{1}{4}\ket{-}\bra{-} \}$ are realized. 
In a similar way, for the six-outcome POVM on the plane $\hat F^6_k$, a BS with transmissivity $\frac{2}{3}$ followed by a BS with transmissivity $\frac{1}{2}$, create three different optical paths where the probability of detecting a photon is $\frac{1}{3}$.
Later, one path is directly measured along with the $\mathds{Z}$ basis with a PBS, implementing the elements $\hat F^6_{1,4}=\frac{1}{6}\ket{H}\bra{H},\frac{1}{6}\ket{V}\bra{V}$. In the second arm an HWP at $\frac{\pi}{12}$ before the PBS implements the elements $\hat F^6_{2,5}=\frac{1}{6}(\frac{\sqrt{3}}{2}\ket{H}+\frac{1}{2}\ket{V})(\frac{\sqrt{3}}{2}\bra{H}+\frac{1}{2}\bra{V}),\frac{1}{6}(\frac{1}{2}\ket{H}-\frac{\sqrt{3}}{2}\ket{V})(\frac{1}{2}\bra{H}-\frac{\sqrt{3}}{2}\bra{V})$.
Similarly, in the third arm an HWP at $\frac{\pi}{6}$ before the PBS implements the elements:
$\hat F^6_{3,6}=\frac{1}{6}(\frac{1}{2}\ket{H}+\frac{\sqrt{3}}{2}\ket{V})(\frac{1}{2}\bra{H}+\frac{\sqrt{3}}{2}\bra{V}), \frac{1}{6}(\frac{\sqrt{3}}{2}\ket{H}-\frac{1}{2}\ket{V})(\frac{\sqrt{3}}{2}\bra{H}-\frac{1}{2}\bra{V})$.
Finally, the implementation of the six-outcome POVM $\hat{S}^6$ is similar to the previous. One of the HWP is now rotated at $\frac{\pi}{8}$ while the other HWP is substituted with a QWP at $\frac{\pi}{4}$. In this way each arm measures along one of the $\mathds{X},\mathds{Y},\mathds{Z}$ bases, implementing the following POVM elements $\{\hat S^6_k\}=\left\{ \frac{1}{6}\ket{H}\bra{H},\frac{1}{6}\ket{+}\bra{+},\frac{1}{6}\ket{L}\bra{L},\frac{1}{6}\ket{V}\bra{V},\frac{1}{6}\ket{-}\bra{-},\frac{1}{6}\ket{R}\bra{R} \right\}$.

After the polarization measurements, the photons are collected by multimode fibers and detected by Silicon SPAD (Excelitas SPCM-NIR).

The electrical signals generated by the SPADs are registered by a Time-to-Digital Converter (TDC) with a resolution of $81\si{\pico\second}$, that streams the data to a PC.
On the PC, we keep only the timetags that are inside a coincidence window of $1\si{\nano\second}$ between the heralding detector and any other detector.
\section{Results}
In this section we describe the results of our experimental run. 
For each of the four measurement configurations described in the previous section, we prepare four different quantum states $\rho_A$ and we evaluate the corresponding min-entropy $H_{min}(X|E)_a$ in the asymptotic limit. The states are chosen in order to maximize or minimize the min-entropy. However, since the protocol assumes uncharacterized light, we don't use any information about the preparation for the actual estimation of the randomness in the system.

For each run of the protocol, we use the heralded source to prepare the state and we record the number of coincidences between the heralding detector $D_0$ and any other detector $D_1$-$D_N$, associated to a particular POVM element. Then the total number of events per detector $N_{k}$ is directly converted to a probability $p_{k}=\frac{N_{k}}{\sum_i N_{k}}$ of the occurrence of a particular POVM element $\hat F_k$.

For a typical run of the experiment we acquire a total number $N_{tot}$ of $10^7$ coincidence events.
However, since the prepared states are (almost) pure, the finite statistics could lead to non-physical quantum states, similarly to what  happens for quantum state tomography~\cite{DAriano2003,James2001MeasurementQubits}.
To enforce a physical reconstruction, we use the constrained maximum-likelihood estimation technique presented in \cite{Faist2016} to retrieve a physical state $\tilde \rho_A$ compatible with the measured statistics $p_{k}$.
The asymptotic min-entropy $H_{min}(X|E)_a$ is then calculated using Eq. \ref{eq:Pg_teo} (or its general version given in Eq. \ref{eq:pg_general}) for the reconstructed state $\tilde \rho_A$.

The results are shown graphically in Fig. \ref{fig:POVM_scaling}, while the estimated $\tilde \rho_A$ and $H_{min}(X|E)_a$ are reported in the Tables \ref{tab:3POVM}, 
\ref{tab:4POVM}, 
\ref{tab:6POVM} and \ref{tab:6POVM_sic} in Appendix \ref{sec:app_result}.
As we can see the experimental data confirm the expected scalings up to $N=6$, for both the maximum and minimum of the $H_{min}(X|E)$.

While the theoretical lower bound was always achievable experimentally (up to numerical precision), the maximum of the $H_{min}(X|E)$ could not be achieved exactly.
This effect is due to the limited accuracy in the preparation of the $\rho_A$ state and unavoidable dark counts in $\hat F_k$ due to accidental coincidences.
\section{Conclusion}

We have presented a protocol for the generation of random numbers from quantum measurement based on the Source-Device-Independent scenario: no assumptions are included in the source of quantum states, while the measurement device is fully trusted and characterized. We have shown that, the amount of extractable random bits scales up as $\propto \log_2(N)$ when the measurement is performed by a $N$-outcome POVM. Then, an infinite number of random bits can be certified for any dimension of the quantum system to be measured.
We derived an analytical bound for the estimation of the extractable randomness using symmetric POVM on the Bloch sphere.

Our findings were validated experimentally by implementing the POVM in the polarization space of single photons with a simple optical setup.

\begin{acknowledgments}
This work was supported by: ``Fondazione Cassa di Risparmio di Padova e Rovigo'' with the project QUASAR funded within the call ``Ricerca Scientifica di Eccellenza 2018''; MIUR (Italian Minister for Education) under the initiative ``Departments of Excellence'' (Law 232/2016); EU-H2020 program under the Marie Sklodowska Curie action, project QCALL (Grant No. GA 675662).
\end{acknowledgments}

\appendix

\section{Proof of analytic results}
In this section, we discuss and prove the analytical bounds on the $H_{min}(X|E)$ presented in the main text.

\begin{figure}[b]
\includegraphics[width=0.75\linewidth]{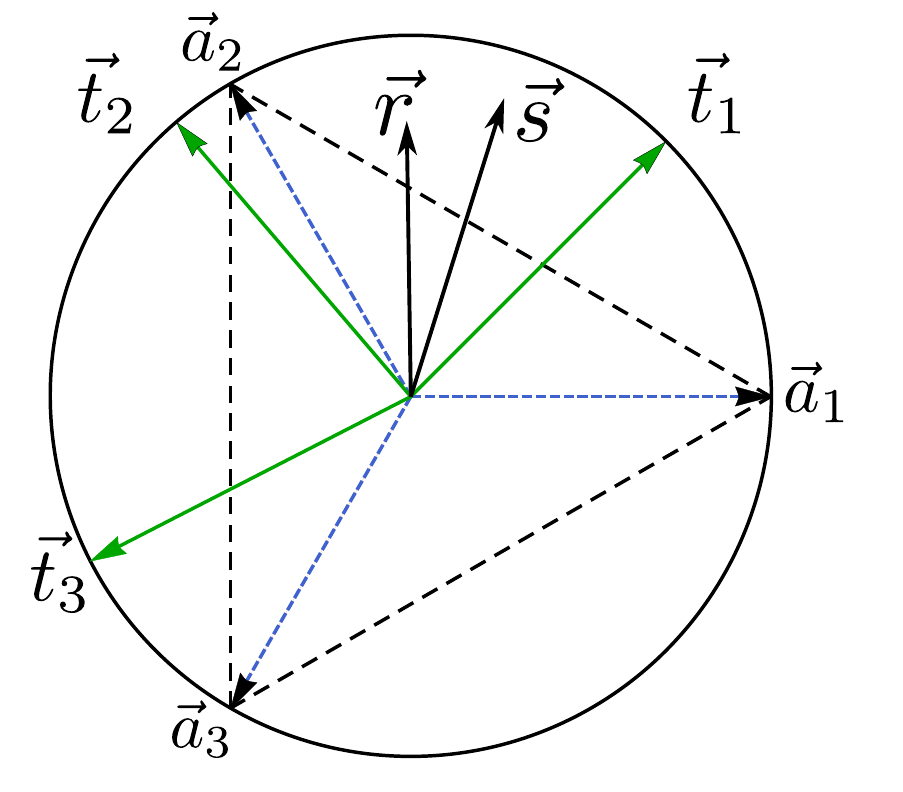}
\caption{ Graphical representation of the 3-outcome POVM.}
\label{fig:Bloch}
\end{figure}

Let's consider the $3$-POVM case defined in \eqref{3Pi}, that can be written as 
\beq
\hat F^3_k =\frac13(\openone+\vec a_k\cdot \vec \sigma)
\eeq
where $\vec \sigma$ are Pauli matrices and the unit vectors $\vec a_k$ represent the vertices of a equilateral triangle are shown in Fig. \ref{fig:Bloch}

Let's suppose that the measurement outcomes are compatible with the state $\vec r$ in Fig. \ref{fig:Bloch}.
First of all, we may notice that the best strategy for Eve is to send an ensemble of pure states. Indeed, if Eve sends a mixed state $\hat \rho'$ with a guessing
probability $P_0=\max_k\text{Tr}[\hat F_k\hat\rho']$, it is always possible to find two pure states $\ket{\psi_1}$ and $\ket{\psi_2}$ such that $\hat\rho'=\lambda\ket{\psi_1}\bra{\psi_1}+(1-\lambda)\ket{\psi_2}\bra{\psi_2}$ and $P'_0=\lambda \max_k\bra{\psi_1}\hat F_k\ket{\psi_1}+(1-\lambda)\max_k\bra{\psi_2}\hat F_k\ket{\psi_2}\geq P_0$.

Thus, the best strategy for Eve is to send three pure states $\hat\tau_k=\frac12(\openone+\vec t_k\cdot \vec \sigma)$ 
with probabilities $p_k$ that satisfy
\beq
\label{r}
p_1\vec t_1+p_2\vec t_2+p_3\vec t_3=\vec r
\eeq
The states $\hat\tau_k$ and probabilities $p_k$ are chosen in order to maximize the guessing probability subjected to the constraint \eqref{r}:
\beq
\label{Pg}
\begin{aligned}
p_g&=\sum_kp_k\text{Tr}[\hat F_k\hat\tau_k]
=\frac13+\frac13\sum p_k\vec t_k\cdot\vec a_k
\end{aligned}
\eeq
If $\vec r$ is within the dashed triangle in Fig. \ref{fig:Bloch}, then the above equation can be solved by choosing $\vec t_k=\vec a_k$ and the appropriate probabilities: this is the best strategy for Eve that achieves the maximum $p_g=\frac23$.

Let's now consider the case in which $\vec r$ is outside the triangle in Fig. \ref{fig:Bloch}. Without losing generality, we may consider that $\vec r$ lies between $\vec a_1$ and $\vec a_2$ we may write the above relations as
\begin{align}
\label{q}
\vec r
&=(1-q) \vec s+q\vec t_3
\notag
\\
\notag
p_g&=\frac{1+q \vec t_3\cdot\vec a_3}{3}+\frac{1-q}3\left[\lambda\vec t_1\cdot\vec a_1+(1-\lambda)\vec t_2\cdot\vec a_2\right]
\end{align}
with $0\leq \lambda,q\leq 1$ and where we have defined the state $\vec s$ by
\beq
\label{s}
\vec s=\lambda \vec t_1+(1-\lambda )\vec t_2
\eeq
First of all, we fix $\vec s$ (and thus $q$ and $\vec t_3$) and try to find the choice for $\vec t_1$, $\vec t_2$ and $\lambda$ that
maximizes $p_g$.
We can consider $\vec t_1$ as variable, while $\vec t_2$ and $\lambda$ should be derived from \eqref{s}.
Indeed, we can find $\lambda$ by squaring the relation $(1-\lambda )\vec t_2=\vec s-\lambda \vec t_1$ and by remembering that $|\vec t_k|=1$. Then, by defining $s=|\vec s|$ we have:
\beq
\label{lambda}
\lambda=\frac{1-s^2}{2-2\vec s\cdot\vec t_1}
\eeq

Due to relation \eqref{s}, the guessing probability can be written as
\beq
\begin{aligned}
p_g&=
\frac{1+q\vec t_3\cdot\vec a_3}{3}+\frac{1-q}3\left[\vec s\cdot\vec a_3+\lambda\vec t_1\cdot(\vec a_1-\vec a_2)\right]
\end{aligned}
\eeq
Since $q$, $\vec t_3$ and $\vec s$ are fixed, Eve should maximize the term
\beq
G=\lambda\vec t_1\cdot(\vec a_1-\vec a_2)
\eeq
Let's define two orthonormal vectors 
$\vec u_k=\vec a_{k}+\vec a_{k+1}$ and
$\vec u^\perp_k=\frac{\vec a_{k}-\vec a_{k+1}}{\sqrt{3}}$. By using \eqref{lambda}, the $G$ function can be written as
\beq
G=\sqrt{3}\frac{1-s^2}{2-2\vec s\cdot\vec t_1}\vec t_1\cdot\vec u_k^\perp
\eeq

\begin{figure}[t]
\includegraphics[width=0.75\linewidth]{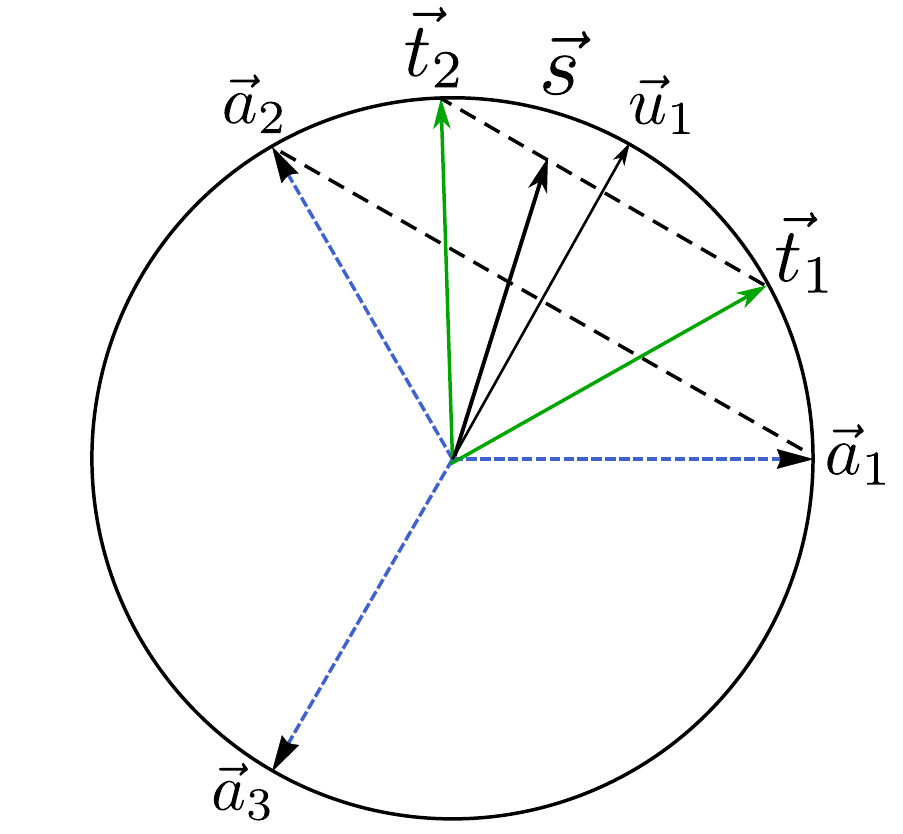}
\caption{Illustration of the states that maximize the function $G$.}
\label{fig:solution}
\end{figure}

Since $(\vec u_k,\vec u_k^\perp)$ is an orthonormal basis, we may
write $\vec s=s(\cos\alpha \vec u_k+\sin\alpha \vec u_k^\perp)$ and $\vec t_1=\cos\theta \vec u_k+\sin\theta \vec u_k^\perp$ such that
\beq
G=\sqrt{3}(1-s^2)\frac{\sin\theta}{2-2s\cos(\theta-\alpha)}
\eeq
Such function is maximized for 
\beq
\cos\theta=\vec t_1\cdot \vec u_k=s\cos\alpha=\vec s\cdot \vec u_k\geq \vec r\cdot \vec u_k
\eeq
corresponding to the situation illustrated in Fig. \ref{fig:solution}.

We may note that the states $\vec t_1$ and $\vec t_2$ are at the same angle with respect to $\vec a_1$ and $\vec a_2$, namely $\vec t_1\cdot \vec a_1=\vec t_2\cdot \vec a_2=\frac{1}2(\vec s\cdot\vec u_1+ \sqrt{3-3(\vec s\cdot\vec u_1)^2})$.
Since $\vec u_1=-\vec a_3$ and $qt_3=\vec r-(1-q)\vec s$, the guessing probability can now be written as
\beq
\begin{aligned}
p_g&=\frac{1-\vec r\cdot\vec u_1}3+\frac{1-q}6 (3\vec s\cdot\vec u_1+
\sqrt{3-3(\vec s\cdot\vec u_1)^2})
\end{aligned}
\eeq

that is maximized for 
\beq
q=0\quad\Rightarrow\quad \vec s=\vec r
\eeq
so the $\vec t_3$ state is never used by Eve.
The guessing probability thus becomes
\beq
\begin{aligned}
p_g
&=\frac{1}3 (1+\frac12\vec r\cdot\vec u_1+\frac{\sqrt 3}{2}
\sqrt{1-(\vec r\cdot\vec u_1)^2})
\end{aligned}
\eeq

\section{$N$ equispaced POVMs}
\label{app:nequi}
Let's consider $N$ equispaced POVMs on a plane of the Bloch sphere with coordinate $(x,y)$.
The POVMs are given by
\beq
\hat F^N_k =\frac1N (\openone+\vec a_k\cdot \vec \sigma)\,,\qquad k=1,\cdots, N
\eeq
with
\beq
\vec a_k=(\cos(\frac{2k\pi}{N}), \sin(\frac{2k\pi}{N}))
\eeq

The vectors $\vec a_k$ identify the vertices of a regular polygon with $N$ edges. We note that 
\beq
\vec a_k+\vec a_{k+1}=2\cos\frac{\pi}{N}\, \vec u_k\,,\qquad
\eeq
where $\vec u_k$ are the unit vectors orthogonal to the polygon edges:
\beq
\vec u_k= (\cos[\frac{(2k+1)\pi}{N}], \sin[\frac{(2k+1)\pi}{N}])
\eeq

The measurement outcome of the POVM identify a state on the $(x,y)$ plane defined by  $\vec r$ (i.e. $\hat\rho=\frac12(1+\vec r\cdot\vec \sigma)$).
The vector $\vec r$ is inside the polygon if and only if
\beq
\vec r\cdot \vec u_k\leq \cos\frac{\pi}{N}\,,\qquad\forall k
\eeq

When the vector $\vec r$ is outside the polygon there is one (and only one) $k$ (say $k^*$) such that $\vec r\cdot \vec u_{k^*}>\cos\frac{\pi}{N}$.

Similarly to the 3-output case, the best strategy for Eve is to send $N$ pure states $\hat\tau_k=\frac12(\openone+\vec t_k\cdot \vec \sigma)$ with probabilities $p_k$ that satisfy
\beq
\label{r_N}
\sum_k p_k\vec t_k=\vec r
\eeq
The states $\hat\tau_k$ and probabilities $p_k$ are chosen in order to maximize the guessing probability 
\beq
p_g=\frac1N+\frac1N\sum p_k\vec t_k\cdot\vec a_k
\eeq
subjected
to the constraint \eqref{r_N}.
If $\vec r$ is inside the polygon, than for Eve it is possible to choose $\vec t_k=\vec a_k$ giving the maximal guessing probability $p_g=\frac2N$.
If the state is outside the polygon, it means that $\vec r\cdot \vec u_{k^*}>\cos\frac{\pi}{N}$ for a single $k^*$.
Similar to the three-outcome case, Eve best strategy is to choose only the two vectors $\vec t_{k^*}$ and $\vec t_{k^*+1}$ that are between $\vec a_{k^*}$ and $\vec a_{k^*+1}$ which  satisfy $\vec t_{k^*}\cdot  \vec a_{k^*}=\vec t_{k^*+1}\cdot  \vec a_{k^*+1}$.
The guessing probability thus becomes
\beq
p_g=\frac1N+\frac1N\vec t_{k^*}\cdot  \vec a_{k^*}
\eeq
Since $\vec t_{k^*}\cdot \vec u_{k^*}=\vec r\cdot \vec u_{k^*}$ we have that 
\beq
\vec t_{k^*}\cdot  \vec a_{k^*}=\vec r\cdot \vec u_{k^*}\cos\frac{\pi}{N}+\sqrt{1-(\vec r\cdot \vec u_{k^*})^2}\sin\frac{\pi}{N}
\eeq

In general then we may write
\beq
\begin{aligned}
p_g&=\frac1N+\frac1N\sum_k
f_N(\vec r\cdot \vec u_k,\frac{\pi }{N})\,
\theta(\vec r\cdot \vec u_k-\cos\frac{\pi }{N})
\end{aligned}
\eeq
where
\beq
f_N(x,\alpha)=
x\cos\alpha+\sqrt{1-x^2}\sin\alpha
\eeq
and
$\theta(x)$ is the Heaviside function.

We note that the angle $2\alpha=\frac{2\pi}{N}$ represents the angle between two consecutive polygon vertices, while $\alpha$ is the angle between $\vec a_k$ and $\vec u_k$.

The above relation can be generalized when the $\vec a_k$ identify the vertices of a Platonic solid. In this case the relation can be generalized to 
\beq
\begin{aligned}
p_g&=
\frac1N+\frac1N\sum_k
f_N(\vec r\cdot \vec u_k,\alpha)
\,
\theta(\vec r\cdot \vec u_k-\cos\alpha)
\label{eq:pg_general}
\end{aligned}
\eeq
with $\vec u_k$ the unit vectors that are normal to the edges and $\alpha$
the angle between $\vec u_k$ and one of the adjacent vertices $\vec a_k$.
For the tetrahedron, octahedron, cube, icosahedron and dodecahedron the angles $\alpha$ are 
$\arccos(1/3)$, $\arccos(1/\sqrt{3})$, $\arccos(1/\sqrt{3})$,
$\arccos(\sqrt{\frac{5+2\sqrt{5}}{15}})$ and
$\arccos(\sqrt{\frac{5+2\sqrt{5}}{15}})$
respectively.

\section{Experimental data}
\label{sec:app_result}
In this section we report the numerical data obtained during the experimental run. For each of the measurement configurations described in the main text, we report the prepared state, the state $\tilde\rho_A$ estimated via Maximum Likelihood Estimation (MLE) from the experimental probabilities, the expected conditional min-entropy $H_{min}(X|E)_t$ for the prepared state and the estimated min-entropy $H_{min}(X|E)_a$ relative to $\tilde \rho_A$ .

\vspace{2cm}
\begin{table}[htb]
\centering
\begin{tabular}{@{}|c|c|c|c|@{}}
\hline\hline\vspace{-0.2cm}\\
State & $H_{min}(X|E)_a$ & $H_{min}(X|E)_t$ & MLE fitted $\tilde \rho_A$ \\ 
\hline\vspace{-0.2cm}\\

$\ket{H}$ & 0.969 & 1.000 & $\begin{bmatrix}
  9.996\cdot10^{-1} & -0.01\\
  -0.01 & 4\cdot10^{-3}\\
\end{bmatrix}$ \vspace{0.1cm}\\ 

\hline\vspace{-0.25cm}\\

$\ket{V}$ & 0.585 & 0.585 & $\begin{bmatrix}
  0.005 & -0.005\\
  -0.005 & 0.995\\
\end{bmatrix}$ \vspace{0.1cm}\\ 

\hline\vspace{-0.25cm}\\
$\ket{+}$  & 0.687 & 0.685 & $\begin{bmatrix}
  0.460 & 0.477\\
  0.477 & 0.540\\
\end{bmatrix}$ \vspace{0.1cm}\\ 

\hline\vspace{-0.25cm}\\
$\ket{L}$  & 0.585 & 0.585 & $\begin{bmatrix}
  0.483 & -0.008\\
  -0.008 & 0.517\\
\end{bmatrix}$ \vspace{0.1cm}\\ 

\hline\hline
\end{tabular}
\caption{Results for the equiangular $3$ outcome POVM on the $XZ$ plane of the Bloch sphere. }
\label{tab:3POVM}
\end{table}

\begin{table}[htb]
\centering
\begin{tabular}{@{}|c|c|c|c|@{}}
\hline\hline\vspace{-0.2cm}\\
State & $H_{min}(X|E)_a$ & $H_{min}(X|E)_t$ & MLE fitted $\hat \rho_A$ \\ 

\hline\vspace{-0.25cm}\\
$\ket{H}$  & 1.000 & 1.000 & $\begin{bmatrix}
  0.998 & 0.001\\
  0.001 & 0.002\\
\end{bmatrix}$  \vspace{0.1cm}\\ 
\hline\vspace{-0.25cm}\\

$\ket{+}$  & 1.000 & 1.000 & $\begin{bmatrix}
  0.502 & -0.499\\
  -0.499 & 0.498\\
\end{bmatrix}$\vspace{0.1cm}\\ 

\hline\vspace{-0.25cm}\\
$\ket{L}$  & 1.000 & 1.000 & $\begin{bmatrix}
  0.499 & -0.005\\
  -0.005 & 0.501\\
\end{bmatrix}$ \vspace{0.1cm}\\ 

\hline\vspace{-0.25cm}\\
$\ket{\frac{\pi}{8}}$  & 1.178 & 1.228 & $\begin{bmatrix}
  0.852 & -0.352\\
  -0.352 & 0.148\\
\end{bmatrix}$ \\ 
\hline\hline
\end{tabular}
\caption{Results for the equiangular $4$ outcome POVM on the $XZ$ plane of the Bloch sphere.
The state $\ket{\frac{\pi}{8}}$ is rotated by ${\frac{\pi}{8}}$ in the $XZ$ plane respect to $\ket{H}$.}
\label{tab:4POVM}
\end{table}

\begin{table}[htb]
\centering
\begin{tabular}{@{}|c|c|c|c|@{}}
\hline\hline\vspace{-0.2cm}\\

State &  $H_{min}(X|E)_a$ & $H_{min}(X|E)_t$ & MLE fitted $\hat \rho_A$ \\ \hline\vspace{-0.2cm}\\

$\ket{H}$ & 1.585 & 1.585 & $\begin{bmatrix}
  0.997  & 0\\
  0 & 0.003\\
\end{bmatrix}$ \vspace{0.1cm}\\ 
\hline\vspace{-0.25cm}\\
$\ket{\frac{\pi}{6}}$  & 1.585 & 1.585 & 	
$\begin{bmatrix}
  0.747 & 0.425\\
  0.425 & 0.253\\
\end{bmatrix}$ \vspace{0.1cm}\\ 
\hline\vspace{-0.25cm}\\
$\ket{L}$  & 1.585 & 1.585 & $ \begin{bmatrix}
  0.505 & -0.006\\
  -0.006 & 0.495\\
\end{bmatrix} $ \vspace{0.1cm}\\ 
\hline\vspace{-0.25cm}\\
$\ket{\frac{\pi}{12}}$  & 1.644 & 1.685 & $\begin{bmatrix}
  0.928 & 0.251\\
  0.251 & 0.072\\
\end{bmatrix}$ \\ 
\hline\hline
\end{tabular}
\caption{Results for the equiangular $6$ outcome POVM on the $XZ$ plane of the Bloch sphere. The states $\ket{\frac{\pi}{6}},\ket{\frac{\pi}{12}}$ are rotated by the respective angles in the $XZ$ plane respect to  respect to $\ket{H}$}
\label{tab:6POVM}
\end{table}

\begin{table}[htbt]
\centering
\small\addtolength{\tabcolsep}{-0.8pt}
\begin{tabular}{@{}|c|c|c|c|@{}}
\hline\hline\vspace{-0.2cm}\\

State & $H_{min}(X|E)_a$ & $H_{min}(X|E)_t$ & MLE fitted $\hat \rho_A$ \\ \hline\vspace{-0.2cm}\\

$\ket{H}$ & 1.585 & 1.585 & $\left[\begin{smallmatrix}
  0.997 & 0.006-0.005i\\
  0.006+0.005i & 0.003\\
\end{smallmatrix}\right]$ \vspace{0.1cm}\\ 
\hline\vspace{-0.25cm}\\
$\ket{+}$ & 1.585 & 1.585 & 	
$\left[\begin{smallmatrix}
  0.502 & 0.494+0.002i\\
  0.494-0.002i & 0.498\\
\end{smallmatrix}\right]$ \vspace{0.1cm}\\ 
\hline\vspace{-0.25cm}\\
$\ket{L}$ & 1.585 & 1.585 & $ \left[\begin{smallmatrix}
  0.503 & 0.003+0.494i\\
  0.003-0.494i & 0.497\\
\end{smallmatrix} \right]$ \vspace{0.1cm}\\ 
\hline\vspace{-0.25cm}\\
$\ket{int}$  & 1.923 & 1.924 & $\left[\begin{smallmatrix}
  0.788 & -0.287-0.291i\\
  -0.287+0.291i & 0.212\\
\end{smallmatrix}\right]$ \vspace{0.1cm}\\ 
\hline\hline
\end{tabular}
\caption{Results for the octahedron $6$ outcome POVM.
The state $\ket{int}$ is in between the three states $\ket{H},\ket{+},\ket{L}$. }
\label{tab:6POVM_sic}
\end{table}

\clearpage
\bibliography{references}
\end{document}